\documentclass[aps,prl,twocolumn,groupedaddress]{revtex4}
\usepackage{graphicx,epsfig,subfigure}
\usepackage{amsmath,amssymb}
\usepackage{color}

\newcommand\beq{\begin{equation}}
\newcommand\eeq{\end{equation}}
\newcommand\beqn{\begin{eqnarray}}
\newcommand\eeqn{\end{eqnarray}}
\newcommand\nn{\nonumber}
\newcommand\fc{\frac}
\newcommand\lt{\left}
\newcommand\rt{\right}
\newcommand\pt{\partial}
\newcommand\tx{\text}

\begin{document}


\title{Gravity localization and mass hierarchy in scalar-tensor branes}



\author{Ke Yang\footnote{yangke09@lzu.edu.cn},
        Yu-Xiao Liu\footnote{liuyx@lzu.edu.cn, corresponding author},
        Yuan Zhong\footnote{zhongy2009@lzu.edu.cn},
        Xiao-Long Du\footnote{duxl11@lzu.edu.cn},
        and Shao-Wen Wei\footnote{weishw@lzu.edu.cn}}.
 \affiliation{Institute of Theoretical Physics, Lanzhou University, Lanzhou 730000,
             China}



\begin{abstract}
We consider a braneworld model in the scalar-tensor gravity. In order to solve the gauge hierarchy problem in this model, our world should be confined on the positive tension brane rather than on the negative one. This is crucial to reproduce a correct Friedmann-like equation on the brane. Interestingly, it is found that the spacing of mass spectrum in this scenario is very tiny, but the light gravitons cannot be observed individually in colliders because of their sufficiently weak interaction with matter fields on the visible brane.
\end{abstract}









\maketitle



\section{I. Introduction}

Motivated by string/M theory, there has been increasing interest in the braneworld scenario during recent years (for an introduction, see, e.g., Refs. \cite{Rubakov2001,Csaki2004,Dzhunushaliev2010} and references therein). In this scenario, our world is trapped in a four-dimensional (4D) submanifold (called brane) embedded in a fundamental multidimensional spacetime (called bulk). This scenario provides a mechanism that could possibly solve some disturbing problems of high-energy physics, such as the gauge hierarchy problem (the problem of why the electroweak scale $M_{\text{EW}}\approx 1$TeV is so different from the Planck scale $M_{\text{Pl}}\approx 10^{16} $TeV) and the cosmological constant problem \cite{Arkani-Hamed1998,Antoniadis1998,Randall1999,Randall1999a}. A famous theory in the braneworld scenario is the Randall-Sundrum (RS) model proposed in 1999 \cite{Randall1999,Randall1999a}. In the RS1 model \cite{Randall1999}, two 3-branes are located at the boundaries of a compact extra dimension with the topology $S^1/Z_2$.
We live on the negative tension brane (called visible brane) and the spin-2 gravitons localize on the positive tension brane (called hidden brane).
The gauge hierarchy problem is solved by introducing an exponential warp factor $e^{-kr_c\phi}$ to warp the extra dimension, and this warped extra dimension generates an exponential hierarchy $e^{-kr_c\pi}\simeq10^{-16}$ to warp down the Planck scale to the electroweak scale on our visible brane. When the radius of $S^1$ approaches infinite, the RS1 model transforms into the RS2 model \cite{Randall1999a}.

Recently, some braneworld models in the scalar-tensor theory have been intensively discussed. The scalar-tensor gravity generalizes the Brans-Dicke theory \cite{Brans1961} and could be traced back to the low-energy effective theory of string/M theory, where a dilaton filed is nonminimally coupled to the gravity. In Ref. \cite{Antoniadis2012}, an interesting linear dilaton model arising from little string theory was proposed. Setting TeV as the fundamental scale, there is a Kaluza-Klein (KK) tower with a TeV mass gap followed by a near continuum of narrow resonances separated from each other by about 30 GeV. In Refs. \cite{Bogdanos2006,Guo2012,Liu2012}, some thick braneworld models were discussed. In general relativity, the profile of the massless graviton is determined by the warp factor only, while it is determined by both the warp factor and the scalar field in the scalar-tensor theory \cite{Guo2012,Liu2012,Farakos2007}. For a weak 4D gravity originating from a small overlap of the massless graviton and the matter fields on the brane, by changing the profile of the massless graviton, one may realize moving our world to the positive tension brane in the scalar-tensor frame. The motivations are as follows: (a) the universe confined on the negative tension brane would lead to a ``wrong-signed" Friedmann-like equation, which causes a severe cosmological problem in the RS1 model. However, this problem can be avoided if our universe is confined on the positive tension brane \cite{Csaki1999,Cline1999,Shiromizu2000}. (b) There are positive energy objects, such as D-branes and NS-branes, on which the Standard Model matter fields can be localized \cite{Lykken2000}.

Therefore, in this work we would like to investigate a simple generation of the RS1 model in the scalar-tensor gravity to move our world to the positive tension brane but keep the advantage of the RS1 model for solving the hierarchy problem.

Here, it is also interesting to note that a model involving only positive tension branes was proposed in Ref. \cite{Lykken2000} by introducing a large tension ``Planck brane" located at the origin of an infinite extra dimension and a small tension probe-brane located at $y_0$. With the distance $y_0$ satisfying $e^{-ky_0}\simeq10^{-16}$, the model can solve the hierarchy problem and recover a consistent low-energy 4D gravity \cite{Lykken2000}.

\section{II. The model}

We start with a 5D action in which gravity is nonminimally coupled to a dilaton field
\beq
S_5=\fc{M^3_{*}}{2}\int{d^5x\sqrt{|g|}e^{k\phi}\lt[R-(3+4k)(\pt\phi)^2\rt]},\label{Action}
\eeq
where $M_{*}$ is the 5D scale of gravity.
Since the scalar-tensor theory has a close relation with the Weyl geometry \cite{Romero2012a}, the action (\ref{Action}) can be originated from a simple Weyl action $S_5^W=\fc{M_*^3}{2}\int_{M^W_5}{d^5x\sqrt{|g|}e^{k\phi}\mathcal{R}}$, where the scalar curvature $\mathcal{R}$ is constructed by the Weylian connection $\Gamma_{MN}^P=\{ _{MN}^P \}-\frac{1}{2}(\phi_{,M} \delta_N^P  + \phi_{,N}\delta _M^P-g_{MN}\phi ^{,P} )$, with $\{ _{MN}^P \}$ the Christoffel symbol. So the scalar could be regarded as a geometric field that provides the ``material" to build the brane configuration and does not couple to the ordinary matter fields \cite{Gogberashvili2010a}.

The metric ansatz is given by
\beq
ds^2_5=a^2(z)(\eta_{\mu\nu} dx^\mu dx^\nu + dz^2), \label{C_Metric}
\eeq
where the conformal coordinate $z\in[-z_b,z_b]$ denotes an $S^1/Z_2$ orbifold extra dimension. It relates to the usual nonconformal metric \cite{Randall1999} with a coordinate transformation $dy=a(z)dz$. In order to be consistent with the 4D Poincar$\acute{\text{e}}$ invariance of the metric, we assume that the scalar field depends on the extra dimension only.
Thus the field equations are read as
\begin{eqnarray}
    k \phi''+\big(k^2 +2k+\fc{3}{2}\big)\phi'^2+2k\fc{a'}{a}\phi'+3\fc{a''}{a}=0, \label{Metric_EOM_2_1}\\
    \big(\phi'-2\fc{a'}{a}\big)\big[(4k +3)\phi'+6\fc{a'}{a}\big]=0,\label{Metric_EOM_2_2}\\
    (4k +3)\phi''+\big(2k^2+\fc{3}{2}k\big)\phi'^2+3(4k+3) \fc{a'}{a}\phi'\nn\\
    -2k \big(2\fc{a''}{a}+\fc{a'^2}{a^2}\big)=0,\label{Metric_EOM_2_3}
\end{eqnarray}
where the prime denotes the derivative with respect to the coordinate $z$. From Eq. (\ref{Metric_EOM_2_2}), we easily get
\beqn
\fc{a'}{a}=\fc{1}{2}\phi',~~~~\text{or}~~~~
\fc{a'}{a}=-\fc{3+4k}{6}\phi'.\label{So_In_EOM}
\eeqn
For $k=-3/2$, these two equations are just equivalent. While for $k=-3/4$, the second equation just gives ${a'}=0$, namely, $a(z)$ is a constant, so we are not interested in this trivial case. After substituting the two equations in (\ref{So_In_EOM}) into Eqs. (\ref{Metric_EOM_2_1}) and (\ref{Metric_EOM_2_3}), we find three independent cases for solving the theory.

{Case 1.}
\begin{eqnarray}
  \phi' = 2\fc{a'}{a}, ~~( k = -\fc{3}{2}).\label{Case1}
\end{eqnarray}
In this case, there is just one constraint on the warp factor and the scalar,
\beq
\phi(z)=2\ln a(z).
\eeq
Thus the brane configuration cannot be uniquely fixed. Because in this case $f(\phi)=e^{-\fc{3}{2}\phi}$, the action (\ref{Action}) is invariant under the rescaling $\bar g_{MN}=e^{-\omega}g_{MN},~~\bar\phi=\phi-\omega$, with $\omega$ an arbitrary smooth function.
The invariance can be easily checked by using the relation between the scalar-tensor action $S_5$ and the weyl action $S^W_5$, and further, making use of the rescaling invariance in Weyl geometry \cite{Romero2012}.
So the breaking of the invariance, i.e., $k \neq -\fc{3}{2}$, such as in case 2 and case 3, is necessary for generating the brane configuration.

{Case 2.}
\begin{eqnarray}
  \phi' &=& 2\fc{a'}{a},~~(k \neq -\fc{3}{2}),\label{Case2_1}\\
  \fc{a''}{a} &=& -\lt(2+2k\rt)\fc{a'^2}{a^2}.\label{Case2_2}
\end{eqnarray}
After redefining the integral parameters to satisfy the $Z_2$-symmetric condition and furthermore to make sure that the null signal takes an infinite amount of time to travel from $z_b$ to $z=0$ when $z_b\rightarrow\infty$, as suggested in RS model \cite{Randall1999,Randall1999a}, the solution is given by
\beqn
a(z)&=&(1+\beta|z|)^{\fc{1}{3+2k}},\\
\phi(z)&=&\fc{2}{3+2k}\ln(1+\beta|z|),
\eeqn
where the parameters are set to $\beta>0$, $k<-\fc{3}{2}$.

It is useful in practice to define an effective energy-momentum tensor $T_{MN}$, which is composed of all scalar terms moved to the right-hand side of the Einstein equations. Since the solution is nonsmooth at the two boundaries, the delta functions in the Einstein tensor and energy-momentum tensor compensate for each other. And these delta functions in effective energy-momentum tensor suggest a thin brane solution. The brane configuration could be more easily seen from the effective energy density $\rho$ referring to static observers, which is defined as $\rho=T_{MN}U^M U^N=-T^0_0$,
\beq
\rho=
\fc{6 (1 + k) \beta^2}{(3+2k)^2(1+\beta |z|)^{\fc{8+4k}{3+2k}}}+\fc{4 k \beta[\delta(z)-\delta(z-z_b)]}{(3+2k)(1+\beta|z|)^{\fc{5+2k}{3+2k}}}. \label{Energy_Density_Case2}
\eeq
It clearly shows that there are two thin branes located at the boundaries $z=0$ and $z=z_b$, respectively.

{Case 3.}
\begin{eqnarray}
\fc{a'}{a} &=& -\fc{3+4k}{6}\phi',~~(k \neq -\fc{3}{2}~~\text{and}~~k\neq-\fc{3}{4}),\label{Case3_2}\\
\phi'' &=& \lt(\fc{3}{2}+k\rt)\phi'^2. \label{Case3_3}
\end{eqnarray}
After redefining the integral parameters, we get the solution
\begin{eqnarray}
a(z)&=&(1+\beta |z|)^{\fc{3+4k}{9+6k}},\label{Warpfactor_Case3}\\
\phi(z)&=&-\fc{2}{3+2k}\ln(1+\beta|z|),\label{Scalar_Case3}
\end{eqnarray}
where $\beta>0$ and $-3/2<k<-3/4$.
With this solution, the effective energy density is expressed as
\beq
\rho=\fc{2 (3+k)(3+4k) \beta^2}{3 (3+2k)^2(1+\beta |z|)^{\fc{24+20k}{9+6k}}}-\fc{4k \beta[\delta(z)-\delta(z-z_b)]}{(3+2k)(1+\beta|z|)^{\fc{15+14k}{9+6k}}}.\label{Energy_Density_Case3}
\eeq

For $T_{MN}\supset-k g^{55}\phi''g_{\mu\nu}\delta^{\mu}_{M}\delta^{\nu}_{N}\supset-g_{\mu\nu}\delta^{\mu}_{M}\delta^{\nu}_{N}[V_{\text{vis}}\delta(z)+V_{\text{hid}}\delta(z-z_b)]$, where $V_{\text{vis}}$ ($V_{\text{hid}}$) plays the role of the effective brane tension, we have $\rho=-T^0_0\supset V_{\text{vis}}\delta(z)+V_{\text{hid}}\delta(z-z_b)$. It means that the prefactors of delta functions in the energy densities of the two cases are nothing but the effective brane tensions. These tensions compensate for the effects produced by the bulk component and hence ensure the existence of 4D flat branes. As shown in Eqs. (\ref{Energy_Density_Case2}) and (\ref{Energy_Density_Case3}), there is a positive tension brane at the origin and a negative one at the boundary $z_b$. So the brane configurations of both cases are similar to that of the RS1 model. However, as we will see in the next section, the localization of massless graviton of these two cases will be quite different from that of the RS1 model, so here we suppose that the Standard Model fields are confined on the positive tension brane at $z=0$, and this is crucial to overcome the severe cosmological problem of the RS1 model.

\section{III. Physical implications}

Here we consider the transverse-traceless tensor fluctuations of the metric (\ref{C_Metric}), which refer to spin-2 gravitons. The perturbed metric is given by
\beq
ds^2=a^2(z)[(\eta_{\mu\nu}+\bar h_{\mu\nu}(x,z))dx^{\mu}dx^{\nu}+dz^2], \label{Perturbational_Metic}
\eeq
where $\bar h_{\mu\nu}$ represent tensor fluctuations and satisfy the transverse-traceless conditions $\eta^{\mu\nu}\bar{h}_{\mu\nu}=\eta^{\lambda\nu}\pt_{\lambda}\bar{h}_{\mu\nu}=0$. With this perturbed metric, the $\mu\nu$ component of the linear perturbed Einstein equations simply gives
\beq
\bar{h}''_{\mu\nu}+3\fc{a'}{a}\bar{h}_{\mu\nu}'+k\phi'\bar{h}'_{\mu\nu}+\Box^{(4)}{\bar{h}_{\mu\nu}}=0.\label{Fluctuation_Eq}
\eeq
Furthermore, we decompose $\bar{h}_{\mu\nu}$ as the form
\beq
\bar{h}_{\mu\nu}(x,z)=\varepsilon_{\mu\nu}(x)A^{-\fc{3}{2}}(z)\Psi(z)\label{Decomposition},
\eeq
where the function $A(z)$ is defined as $A(z)=a(z)e^{{k}\phi/{3}}$, and with the solutions in case 2 and case 3, the expressions of $A(z)$ are found to be the same in both cases,
\beq
A(z)=(1+\beta |z|)^{\fc{1}{3}}.
\eeq
The 4D mass $m$ of a KK excitation is introduced by the 4D Klein-Gordon equation
$\Box^{(4)}{\varepsilon_{\mu\nu}(x)}=m^2 \varepsilon_{\mu\nu}(x)$.
Then a Schr$\ddot{\tx{o}}$dinger-like equation is obtained from Eq. (\ref{Fluctuation_Eq}) as
\beq
-\Psi''(z)+\left(\fc{3}{2}\fc{A''}{A}+\fc{3}{4}\fc{A'^2}{A^2}\right)\Psi(z)=m^2\Psi(z).\label{Schrodinger_Eq}
\eeq
The eigenvalue $m$ in this Schr$\ddot{\tx{o}}$dinger-like equation parametrizes the spectrum of the 4D graviton masses. Setting $m=0$ in Eq. (\ref{Schrodinger_Eq}), one can easily get the normalized zero mode
\beq
\Psi_0(z)=\fc{A^{\fc{3}{2}}(z)}{N_0}=\fc{\lt(1+\beta|z|\rt)^{1/2}}{\sqrt{2z_b+\beta z_b^2}},\label{Gravitational_zero_modes}
\eeq
and it is determined by both the warp factor and the scalar field. Here the zero mode is localized on the negative tension brane at $z_b$ instead of on the positive tension one at the origin, although the warp factor decreases towards $z_b$ like the RS1 model. As a result of a compact extra dimension, the zero mode is normalizable, but it is not normalizable anymore when the size of the extra dimension approaches infinity. Since a normalizable massless mode ensures that an effective 4D gravity recovers on the brane at low-energy scale, compactifying the extra dimension is crucial in our model.

With the Neumann boundary condition $\pt_z \bar{h}_{\mu\nu}(x,z)=0$, which is chosen to be consistent with the $Z_2$-symmetry, the general solution of the Schr$\ddot{\tx{o}}$dinger-like equation between the two boundaries is given by a linear combination of the Bessel functions, i.e.,
\beq
\Psi(z)=\fc{(1+\beta z)^{\fc{1}{2}}}{N}\bigg[\text{J}_{0}\big(m(z+\fc{1}{\beta})\big)+\alpha\text{Y}_{0}\big(m(z+\fc{1}{\beta})\big)\bigg],
\eeq
where $\alpha=-\text{J}_{1}(m/\beta)/\text{Y}_{1}(m/\beta)$ and $N$ is a normalization constant.
The KK spectrum is determined by
\beq
\fc{\text{J}_{1}\big(m_n(z_b+\fc{1}{\beta})\big)}{\text{J}_{1}(m_n/\beta)}
      =\fc{\text{Y}_{1}\big(m_n(z_b+\fc{1}{\beta})\big)}{\text{Y}_{1}(m_n/\beta)}.\label{Mass_Spectrum_Condition}
\eeq

Further, working in the limit of $m_n/\beta\ll1$ and $1+\beta z_b\gg1$, i.e., here we consider the light modes in the long-range case, $\alpha\approx({\pi m^2})/({4\beta^2})\ll1$, then the massive modes are given by
\beq
\Psi_n(z)\approx \fc{(1+\beta z)^{\fc{1}{2}}}{\sqrt{2z_b+\beta z_b^2}}\text{J}_{0}\big(m_n(z+1/\beta)\big),~(n>0).
\label{Gravitational_massive_modes}
\eeq
And in this limit, the spectrum is read as
\beq
m_{n}=\fc{x_n}{z_b+1/\beta} \label{Mass_Spectrum},
\eeq
where $x_n$ satisfies $\text{J}_{1}(x_n)=0$, and $x_1=3.83$, $x_2=7.02$, $x_3=10.17,\cdots$.

The interaction between gravitons and matter fields is achieved by including the action of Standard Model matter fields. When the matter fields are located on the visible brane, the interaction in the 4D effective theory is described by \cite{Davoudiasl2000}
\beq
L_{\text{int}}=\fc{\xi}{2}\tilde T^{\mu\nu}(x)h_{\mu\nu}(x,0), \label{Interaction_G_M}
\eeq
where the parameter $\xi=2/M_{*}^{3/2}$ is chosen to give the 5D field $h_{MN}$ a correct dimension, namely, $h_{MN}\rightarrow\xi h_{MN}$, and $\tilde{T}^{\mu\nu}(x)$ is the symmetric conserved Minkowski space energy-momentum tensor. For $z=0$, $\text{J}_0(m_n/\beta)\approx1$, Eqs. (\ref{Gravitational_zero_modes}) and (\ref{Gravitational_massive_modes}) show that $\Psi_n(x,0)\approx\Psi_0(x,0)$. It means that the profiles of lower KK states are similar to the profile of the massless graviton. So with the decomposition (\ref{Decomposition}), one arrives at \beq
h^{(n)}_{\mu\nu}(x,0)=\fc{\varepsilon^{(n)}_{\mu\nu}(x)}{\sqrt{2z_b+\beta z_b^2}},~(n\geq0),
\eeq
and hence, the interaction Lagrangian (\ref{Interaction_G_M}) simply gives
\beq
L^{(n)}_{\text{int}}=\fc{\xi\tilde T^{\mu\nu}(x)\varepsilon^{(n)}_{\mu\nu}(x)}{2\sqrt{2z_b+\beta z_b^2}}=\tilde\xi\tilde T^{\mu\nu}(x)\varepsilon^{(n)}_{\mu\nu}(x), \label{Coupling_Gravity_Matter}
\eeq
where $\tilde\xi=1/[M_{*}^{3}(2z_b+\beta z_b^2)]^{1/2}$ is the effective coupling constant. It shows that the couplings of both the massless and massive gravitons to matter are of the same order. This is quite different from that of the RS1 model, where the two couplings are of order $1/\text{M}_{\text{Pl}}$ and $1/\text{TeV}$, respectively.

After calculating the contribution of the massless zero mode sector in the action (\ref{Action}), the 4D effective Planck scale $M^2_{\text{Pl}}$ is given by
\beq
M^2_{\text{Pl}}=M_{*}^3\int_{-z_b}^{z_b}{dz A^3(z)}=M^3_{*}\lt(2z_b+\beta z_b^2\rt). \label{Relation_M4_M5}
\eeq

On the other hand, since the Standard Model fields are confined on the brane at the origin, where the warp factor $a(0)=1$, the Higgs field action on the visible brane is just the usual 4D canonically normalized one, and the Higgs vacuum expectation value $v_0$ does not involve the warped extra dimension \cite{Randall1999}. Since the Higgs vacuum expectation value sets all mass parameters, any effective physical mass $m_{\text{obs}}$ on the visible brane is identical to its mass parameter $m_{*}$ in the fundamental theory.

Thus, if we set all the fundamental parameters $M_{*}, \beta, v_0$ to be about the TeV scale, one only requires $\beta z_b\approx10^{16}$ to provide a large twist of the two scales $\beta M^2_{\text{Pl}}\approx10^{32}M^3_{*}$.
Then, the mass spectrum is $m_{n}\approx x_n \cdot10^{-4}$eV. Thus in contrast with that of the RS1 model where the spacing of KK gravitons is of order of the TeV scale, the spacing is quite small here.

It seems that these gravitons are tiny enough to be easily produced in all colliders. Nevertheless, the effective coupling constant of KK gravitons with matter on the visible brane is $\tilde\xi=1/M_{\text{Pl}}$ in this case. It means that the interactions of the massless and massive gravitons with matter fields on the visible brane are both largely suppressed by the 4D gravitational strength $1/M_{\text{Pl}}$. Therefore, these light KK gravitons can certainly not be seen individually. Moreover, following the same spirit of the large extra dimension model \cite{Arkani-Hamed1998,Antoniadis1998}, the cross section for real emission of these KK gravitons is roughly
$\sigma\sim{({E}/{10^{-4}\tx{eV}})}/{M^{2}_{\tx{Pl}}}=10^{-16}E/{(\tx{TeV})^3}$,
where $E$ is the relevant physical energy scale. Therefore, their collective contribution is also small enough to ensure that the model is not contrary to the experimental observations.

\section{IV. Discussion}

In this work, we investigate a generation of the RS1 model in the scalar-tensor theory. In order to solve the gauge hierarchy problem in this model, our world should be confined on the positive tension brane rather than on the negative one. This is crucial to reproduce a correct Friedmann-like equation on the brane. The spacing of the KK tower is found to be very tiny, about $10^{-4}$eV. Nevertheless, it will not cause any unaccepted signal in the infrared observations for the largely suppressed interaction of these KK gravitons with matter.

For the parameter $k=-1$, the action $S_5$ is just the standard bosonic part of the effective string action involving only the metric and the dilaton. Thus the brane solution with $k=-1$ in case 3 could be embedded into a string theory.

A constant bulk energy density is obtained in (\ref{Energy_Density_Case2}) and (\ref{Energy_Density_Case3}) for $k=-2$ and $k=-6/5$, respectively. After utilizing the coordinate transformation $dy=a(z)dz$ to rewrite the solutions in a nonconformal coordinate, it is found that the warp factor is exponential, $a(y)=e^{-\beta y}$, like the RS model,
and the scalar is linear
for these values of $k$. Thus, the bulk is an anti-de Sitter spacetime with the effective cosmological constant $\Lambda=-6\beta^2$. However, for other values, the warp factor is power law and the scalar is logarithmic.

With a conformal transformation $\tilde g=e^{2k\phi/3}g$, the action $S_5$ can be written in the Einstein frame as
$
S_5=\fc{M^3_{*}}{2}\int{d^5x \sqrt{|\tilde g|}[\tilde R-\fc{1}{3}{(3+2k)^2}(\pt\phi)^2]}.
$
Therefore, the scalar does not involve the ghost instability problem.
On the other hand, since the bulk size is not dynamically fixed in our model, the Goldberger-Wise mechanism \cite{Goldberger1999a} should also be further introduced to stabilize the size of the extra dimension, and the scalar field could play the role of the stabilizing field \cite{Antoniadis2012,Cox2012}.

\section{Acknowledgement}

The authors would like to thank Heng Guo and  Feng-Wei Chen  for helpful discussions. This work was supported in part by the National Natural Science Foundation of China (Grant No. 11075065),
the Doctoral Program Foundation of Institutions of Higher Education of China
(Grant No. 20090211110028), the Huo Ying-Dong Education Foundation of
Chinese Ministry of Education (Grant No. 121106). Ke Yang and Yuan Zhong were supported by the Scholarship Award for Excellent Doctoral Student granted by Ministry of Education.



\end{document}